\documentstyle[aps,prb,bezier,multicol,epsf]{revtex}
\begin{document}
\title{
   Hikami boxes and the Sinai billiard.
}
\author{ Daniel~L.~Miller }
\address{
 Dept. of Physics of Complex Systems,\\
 The Weizmann Institute of science,
 Rehovot, 76100 Israel                \\
 e-mail  fndaniil@wicc.weizmann.ac.il
}

\date{\today}

\maketitle

\begin{abstract}
   Diagram, known in theory of the Anderson localization as the Hikami box,
   is computed for the Sinai billiard. This interference effect is mostly
   important for trajectories tangent to the opening of the billiard. This
   diagram is universal at low frequencies, because of the particle number
   conservation law. An independent parameter, which we call phase volume of
   diffraction, determines the corresponding frequency range.  This result
   suggests that level statistics of a generic chaotic system is not
   universal.
\end{abstract}

\begin{multicols}{2}

\narrowtext
\unitlength=1.00mm
\linethickness{0.4pt}


Any function, which describes the physical properties of a single
particle bounded system, is called universal in this work, if it is
a function of either frequency or time and has the only one parameter:
the mean level spacing $\Delta$ divided by $\hbar$.

\section{ Formal derivation. }

Following Ref.\onlinecite{Smilansky-rev95} the quantum evolution of the
particle inside the two-dimensional billiard is convenient to describe by the
scattering matrix $S(\theta,\theta')$. For given energy $E=\hbar^2 k^2/(2m)$
the wave numbers of the incident and scattered wave functions are
$\vec k_i = (k\cos(\theta), k\sin(\theta))$ and $\vec k_f = (k\cos(\theta'),
k\sin(\theta'))$ correspondingly. The wave reflected $n$ times from the
billiard walls is described by the $n$-th power of the matrix $S$.

In the particular case of the Sinai billiard one needs the product of two
matrices $S = S_{\framebox(1.20,1.20){}}^\dagger
S_{\makebox(1.5,0.5)[ct]{\circle{1.20}}}$, where $S_{\framebox(1.20,1.20){}}$
is the scattering matrix of the square, and
$S_{\makebox(1.5,0.5)[ct]{\circle{1.20}}}$ is the scattering matrix of the
circle. A similar decomposition was used for quantization of the Sinai
billiard.\cite{Berry-apr80}

\begin{mathletters}
The direct product of two scattering matrices is a density evolution operator.
The two ways to pair arguments of two matrices give diffusion and
cooperon-like operators:
\begin{eqnarray}
   S_{E+\omega}^n( \theta_1\, \theta_2)
   S^{\dagger n}_E(\theta_3 \, \theta_4)
   &=& \sum_{l_1 l_2}
   e^{i(\theta_1-\theta_4) l_1 - i(\theta_2-\theta_3) l_2 }
\nonumber\\
   &\times&
   {\cal D}_\omega^n (\theta_1, l_1; \theta_2, l_2) \;,
\label{eq:form.1}
\\
   S_{E+\omega}^n( \theta_1\, \theta_2)
   S^{\dagger n}_E(\theta_3 \, \theta_4)
   &=& \sum_{l_1 l_2}
   e^{i(\theta_1-\theta_3+\pi) l_1 - i(\theta_2-\theta_4-\pi) l_2 }
\nonumber\\
   &\times&
   {\cal C}_\omega^n (\theta_1, l_1; \theta_2, l_2)\;,
\label{eq:form.2}
\end{eqnarray}
where $l = [\vec r\times \vec k]_z $  is the angular momentum measured in
units of $\hbar$, the $z$-axis points out of the plane.
\label{eq:form.3}
\end{mathletters}

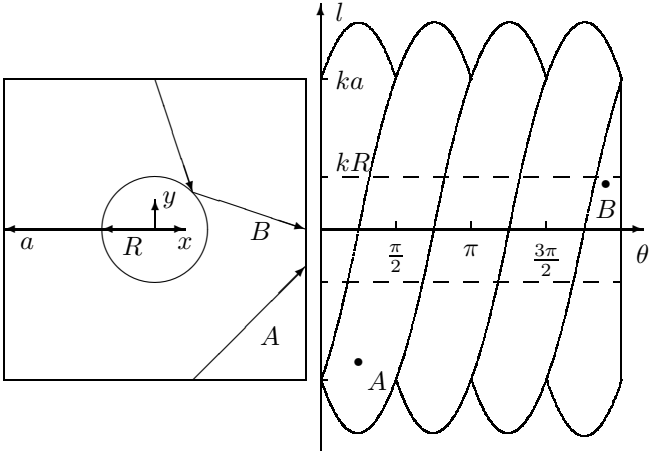
\begin{figure}
\centerline{
\unitlength=1.00mm
\linethickness{0.4pt}
\begin{picture}(84.91,60.00)
\put(20.00,30.00){\circle{14.00}}
\put(0.00,10.00){\framebox(40.00,40.00)[cc]{}}
\bezier{264}(47.00,30.00)(55.00,72.00)(62.00,50.00)
\bezier{264}(57.00,30.00)(65.00,72.00)(72.00,50.00)
\bezier{264}(67.00,30.00)(75.00,72.00)(82.00,50.00)
\bezier{128}(42.00,50.00)(47.00,65.00)(52.00,50.00)
\bezier{84}(77.00,30.00)(80.00,44.00)(82.00,50.00)
\bezier{256}(42.00,10.00)(49.00,-11.00)(57.00,30.00)
\bezier{256}(52.00,10.00)(59.00,-11.00)(67.00,30.00)
\bezier{256}(62.00,10.00)(69.00,-11.00)(77.00,30.00)
\put(82.00,50.00){\line(0,-1){0.08}}
\bezier{128}(72.00,10.00)(77.00,-5.00)(82.00,10.00)
\bezier{84}(42.00,10.00)(44.00,15.00)(47.00,30.00)
\put(42.00,30.00){\vector(1,0){42.91}}
\put(52.00,30.00){\line(0,1){1.00}}
\put(62.00,30.00){\line(0,1){1.00}}
\put(72.00,30.00){\line(0,1){1.00}}
\put(42.00,50.00){\line(1,0){1.00}}
\put(42.00,10.00){\line(1,0){1.00}}
\put(42.00,23.00){\dashbox{2.00}(40.00,14.00)[cc]{}}
\put(25.00,10.00){\vector(1,1){15.00}}
\put(34.00,17.00){\makebox(0,0)[lt]{$A$}}
\put(47.05,12.39){\circle*{1.00}}
\put(48.21,10.99){\makebox(0,0)[lt]{$A$}}
\put(25.00,35.00){\vector(3,-1){15.01}}
\put(20.00,50.00){\vector(1,-3){5.01}}
\put(34.00,31.00){\makebox(0,0)[ct]{$B$}}
\put(79.98,36.04){\circle*{1.00}}
\put(79.98,33.92){\makebox(0,0)[ct]{$B$}}
\put(84.00,28.00){\makebox(0,0)[lt]{$\theta$}}
\put(52.00,28.00){\makebox(0,0)[ct]{$\pi\over 2$}}
\put(62.00,28.00){\makebox(0,0)[ct]{$\pi$}}
\put(72.00,28.00){\makebox(0,0)[ct]{$3\pi\over 2$}}
\put(44.00,59.00){\makebox(0,0)[lc]{$l$}}
\put(44.00,50.00){\makebox(0,0)[lc]{$ka$}}
\put(44.00,39.00){\makebox(0,0)[lc]{$kR$}}
\put(20.00,30.00){\vector(-1,0){7.00}}
\put(13.00,30.00){\vector(-1,0){13.00}}
\put(17.00,29.00){\makebox(0,0)[ct]{$R$}}
\put(3.00,29.00){\makebox(0,0)[ct]{$a$}}
\put(42.00,0.00){\vector(0,1){60.00}}
\put(82.00,50.00){\line(0,-1){40.00}}
\put(20.00,30.00){\vector(0,1){4.00}}
\put(20.00,30.00){\vector(1,0){4.00}}
\put(21.00,34.00){\makebox(0,0)[lc]{$y$}}
\put(24.00,29.00){\makebox(0,0)[ct]{$x$}}
\end{picture}
}
\caption{
  Configuration and phase space of the Sinai billiard. Four areas in the
  phase space correspond to the four sets of trajectories hitting four walls
  of the square.  Reflections from walls generate a map. It takes the
  particle from one area of the phase space and put it to other area.
}
\label{fig:sniphase}
\end{figure}

Operator ${\cal D}$ is a one step classical evolution operator. It is well
defined if position of a classical particle on the energy shell of the phase
space $\vec r, \vec k/k$ is described by $l,\theta$. This is the case for the
Sinai billiard, see Fig.~\ref{fig:sniphase}, if we agree to compute $l$ and
$\theta$ for a particle, just before it hits the square. Thus each point in
the phase space corresponds to either straight piece of trajectory, for
example the point A, or trajectory with one reflection from the circle, for
example the point B in Fig.~\ref{fig:sniphase}.

\begin{mathletters}
Each trajectory in the configuration space is a sequence of straight
lines (segments), it becomes a sequence of points in the phase space; the
reflection law generates a map in the phase space
\begin{equation}
     2 = {\cal M}(1)\;,\quad
     1 \equiv (\theta_1, l_1)\;,\quad
     2 \equiv (\theta_2, l_2)\;.
\label{eq:form.4}
\end{equation}
The density evolution operators Eq.~(\ref{eq:form.3}) computed
within semiclassical\cite{Smilansky-rev95} and
diagonal\cite{Berry-feb85} approximations are ${\cal C}\approx C$,
and ${\cal D}\approx D$, where
\begin{equation}
  C_\omega^n(1;2) = D_\omega^n(1;2)
  = 2\pi e^{i\omega t_{1,2}}
  \delta( 2 - {\cal M}^n(1))
  \;,
\label{eq:form.6}
\end{equation}
${\cal M}^n$ means $n$ iterations of the map $M$, and $t_{1,2}$ is the
time of flight along the trajectory.
\label{eq:form.6.02} \end{mathletters}

The diagonal approximation is justified when classical actions of different
trajectories are not correlated. The action of the trajectory is proportional
to its length. The only scale  of the length-length correlation function is
the system size\cite{Miller-apr98}, and the scale of the
action correlations is, therefore, $\hbar N_H$, where $N_H$ is the number of
open channels in the system, it is the effective dimensionality of the
scattering matrix\cite{Smilansky-rev95}, and it is the analog of the
Heisenberg time for maps. This argument provides us the condition
\begin{equation}
   n \ll N_H \quad\text{or}\quad \omega \gg \Delta/\hbar\;,
\label{eq:form.5.01}
\end{equation}
where $\Delta$ is the mean level spacing.

Under the condition Eq.~(\ref{eq:form.5.01}) the density evolution operator
must preserve the invariant measure\cite{Schuster-book84}, and therefore
$\sum_n \text{tr} { \cal D}_\omega^n$ must have single pole at $\omega=0$ or
alternatively
\begin{equation}
   \text{ tr} { \cal D}_{\omega=0}^n = 1 \;,\quad
   n\gg n_\ast
\label{eq:form.5.02}
\end{equation}
This sum rule holds for the Frobenius - Perron operator\cite{Schuster-book84}
Eq.~(\ref{eq:form.6}). In this case it is known also as the Hannay - Ozorio de
Almeida sum rule\cite{Hannay-OzorioDeAlmeida}, and $n_\ast$ characterizes
decay of Frobenius - Perron modes.

Dashed lines in Fig.~\ref{fig:sniphase} mark the parts of the phase space
where the semiclassical approximation fails because of diffraction. The phase 
volume of diffraction is much smaller than the phase volume of the system.
Therefore ${\cal D} \approx D + \delta D$, and the first order correction to 
the evolution operator is 
\begin{eqnarray}
  \delta{ D}^{n}_\omega(1;2)
  & = &
  \int d3 d4
  \sum_{n'=1}^{n-1}
   D^{n'}_\omega(1;3)
\nonumber\\
  & \times &
  \Bigl[ F_{D}(3;4) - \delta(3;4)\Bigr]
  D^{n-n'}_\omega(3;2)
\label{eq:form.6.01}
\\
  F_{D}(1;2) & = &
  \delta_{l_1 l_2} f_D(l_1, \theta_1-\theta_2)
\label{eq:form.6.04}
\\
   f_{D}(l, \theta) & = &
   \int d l'
   S_{l + l'/2} S^\ast_{l - l'/2} e^{il'\theta}
   \;,
\label{eq:form.6.03}
\end{eqnarray}
where $\delta(1;2) = 2\pi\delta_{l_1 l_2} \delta(\theta_1-\theta_2)$,
$3\equiv(\theta_3, l_3)$,  $4\equiv(\theta_4, l_4)$, $\int d3 = \sum_{l_3}
\int{d\theta_3\over 2\pi}$, $S_l = - H_l^{(1)}(kR) / H_l^{(2)}(kR) $ is the
scattering matrix of a circle. The diffraction coefficient $f_D(l,\theta)$
for $kR - \alpha' < l < kR + \alpha''$, where
$\alpha'\sim\alpha''\sim(kR)^{1/3}$, may be approximated in a number of
ways\cite{Levy-Keller-59}, see Fig.~\ref{fig:kernel}. Outside this interval,
for $l < kR-\alpha'$, it must reproduce the map $f_D(l,\theta) = 2\pi\delta(
\theta - \text{arccos} ({ l\over kR }) )$, and for $l > kR + \alpha''$ one
has $f_D(\theta) = 2\pi\delta(\theta)$.

In order to compute the so-called interference correction one should examine
all possible pairing of arguments in the product of $S$ matrices. The further
computation is eventually the same as in Ref.\onlinecite{Aleiner-Larkin-96}
and the correction mixing two diffusion and one cooperon like operators is
\begin{eqnarray}
   \lefteqn{ \delta D^{n}(1,2)
   = \int d3\,d4\,}
\label{eq:form.10}
\\
   &\times&
   \biggl\{
      D (1; 3)
      D ( \bar 4; 2)
      \Bigl[
         F_I(3;4) C ( 4; \bar 3)
         +
         F_I^\ast(3; 4)  C ( 3; \bar 4)
      \Bigr]
\nonumber\\
   &+&
   D (1; 3) D (\bar 3; 2)
   \bigl[
      F_D(3;4) C ( 4; \bar 3)
      + F_D(\bar 4; \bar 3) C ( 3 ; \bar 4)
   \Bigr]
\nonumber\\
    &+&
     D (1; 3) D (\bar 4; 2)
   \Bigl[
      F_D(3;4) C ( 4; \bar 4)
      + F_D(\bar 3; \bar 4) C ( 3 ; \bar 3)
   \Bigr]
   \biggr\}
\nonumber
\end{eqnarray}
where , $\bar 3 \equiv (\theta_3+\pi,-l_3)$, $\bar 4 \equiv ( \theta_4+\pi,
-l_4)$. The right hand side of Eq.~(\ref{eq:form.10}) should be understood as
a sum over all possible powers $n_1+n_2 + n_3 =n$ of the density evolution
operators, i.e.  $\delta D^{n}\propto  D^{n_1} C^{n_2} D^{n_3}$.  The
diffraction kernels are defined in the vicinity of tangency to the
cylindrical mirror:
\begin{eqnarray}
   F_I(l_1,\theta_1;l_2,\theta_2)
   &=& \delta_{l_1 l_2}
   f_{I}(l_1, \theta_1 - \theta_2)
\label{eq:form.19}
\\
   f_{I}(l, \theta ) &=&
   \int d l'
   S_{l + l'/2} S_{-l + l'/2} e^{il'\theta}\;,
\label{eq:form.20}
\end{eqnarray}
where $S_l$ is the same as in Eq.~(\ref{eq:form.6.03}), and 2Re$f_I$ for $kR
= 50$ is shown in Fig.~\ref{fig:kernel}. In order to avoid difficulties with
self-tracing trajectories we put $F_I(1;2)=F_D(1;2) = 0$ for $l_1<kR-\alpha'$
and $l_1 > kR + \alpha'' $.

The choice of the constants $\alpha', \alpha''$ is restricted by
the particle number conservation law
\begin{equation}
  \int d3\,d4\,
  \Bigl[ F_{I}(3;4) + F_{I}^\ast(3;4) + F_{D}(3;4) \Bigr]
  = 0\;.
\label{eq:form.30}
\end{equation}
Indeed, the interference correction Eq.~(\ref{eq:form.10}) has a form
of the interaction of the Frobenius - Perron modes. Therefore, the sum over
all diffraction and interference corrections shifts the $\omega=0$ pole of
the evolution operator.  In order to avoid the contradiction with the
particle number conservation law in the form of Eq.~(\ref{eq:form.5.02}) the 
shift must be less or of the order of the mean level spacing $\Delta$. ( In 
this case the shift should be neglected because of the condition 
Eq.~(\ref{eq:form.5.01}).)  The shift of the pole depends on the
integral Eq.~(\ref{eq:form.30}). After all this integral must be so small,
that we put it equal to zero.

\begin{figure}
\unitlength=1mm
\begin{picture}(83,30)
\put(7,1){\makebox(0,0)[lb]{$l$}}
\put(30,1){\makebox(0,0)[rb]{$\theta$}}
\put(0,20){\makebox(0,0)[rb]{$f_D$}}
\put(50,1){\makebox(0,0)[lb]{$l$}}
\put(73,1){\makebox(0,0)[rb]{$\theta$}}
\put(43,20){\makebox(0,0)[rb]{$f_I$}}
\end{picture}
\caption{
    The numerical calculation of the diffraction and interference
    kernels.
}
\label{fig:kernel}
\end{figure}

The interference correction Eq.~(\ref{eq:form.10}) contains a small 
parameter: the probability for a given trajectory to visit the diffraction 
region twice, with the same value of angular momentum. The probability to 
find two points of trajectory of length $n$ in the certain part of the phase 
space growth like $n^2$ and this is expected behavior of all terms in the 
right hand side of Eqs.~(\ref{eq:form.10}). Because of Eq.~(\ref{eq:form.30}) 
these terms have different signs and the overall result behaves like $n$. 
There is a similar effect in theory of disordered metals. The interference 
correction to the density evolution operator consists from three evolution 
operators connected by the kernel.  This kernel is called the Hikami 
box\cite{Hikami-81}; it is proportional to the small factor $\omega\tau$ 
because of the well known cancelation\cite{Gorkov-LK-79}.


\section{ Level statistics }
\label{sec:lvl}

The form-factor of the two-point correlation function of energy levels
is expressed in terms of scattering matrix traces\cite{Smilansky-rev89}
\begin{eqnarray}
   K(\tau) &=& {1\over N_H} |\text{tr} S^{n} |^2
   \approx {n\over N_H}\text{tr}\left[
      {\cal D}^{n}_{\omega = 0}
      + {\cal C}^{n}_{\omega = 0}
   \right]\;,
\nonumber\\
   \tau &\equiv& n/N_H
\label{lvl.01}
\end{eqnarray}
In the first order in $\tau$ one obtains $K(\tau)=2\tau$, because of 
Eq.~(\ref{eq:form.5.02}); this universal result is a consequence of the 
ergodicity\cite{Berry-87}.

We will go beyond this approximation by taking into account the interference
correction to the both diffusion and cooperon evolution operators. We should 
introduce the third kind of evolution operators, which would take into 
account exactly the diffraction corrections Eq.~(\ref{eq:form.6.01}). Let us,
instead, include these corrections into the definition of the operators $D$ 
and $C$. Then the next order in $\tau$ correction to the form-factor becomes 
\begin{eqnarray}
  \lefteqn{ \delta K(\tau)
  =
   \tau n
   \int d3\,d4
   \sum_{n'}
   C^{n'}_{\omega = 0} ( 3; \bar 4)
   }
\label{eq:lvl.03} \\
  &\times& \biggl\{
   D^{n-n'}_{\omega = 0} ( \bar 4; 3)\,2\text{Re}F_I(3;4)\,
   +
   D^{n-n'}_{\omega = 0} ( \bar 3; 3)\,F_D(3;4)
   \biggr\}
\nonumber
\end{eqnarray}
The first term in the braces is the sum over periodic orbits of length $n$
visiting twice the diffraction region with the same value of angular 
momentum. The probability to find such a trajectory is 
\begin{equation}
   {n(n-1)\over 2} {\alpha \over \Omega}{1\over N_H}\;,
   \label{eq:lvl.04}
\end{equation}
where $\alpha = \int d1\,d2\,F_D(1,2) \approx (kR)^{1/3}$ is the effective
phase volume of the diffraction region, and the $\Omega$ is the volume of the
phase space; it is just twice the perimeter of the billiard $\Omega = 8ka$.

The second term in the braces in Eq.~(\ref{eq:lvl.03}) is the sum over
periodic orbits which has a self-tracing piece. The probability to find such
a trajectory is
\begin{equation}
   {n(n-1)\over 2} {\alpha \over \Omega}{1\over N_H}
   \left( 1 - e^{-j\alpha/\Omega} \right)\;,
   \label{eq:lvl.05}
\end{equation}
where $j$ is the length of the self-tracing part of the trajectory. The first
term in the braces in Eq.~(\ref{eq:lvl.03}) partially cancels the second 
term, because of Eq.~(\ref{eq:form.30}).  Averaging over $j$ and assuming 
$n\gg 1$ we obtain 
\begin{equation}
  K(\tau) = 2\tau - \tau^2
  \left( 1 - e^{ - {N_H\over 2\Omega} \alpha \tau}\right)
  + O(\tau^3)
\label{eq:lvl.06}
\end{equation}
where the total amount of periodic trajectories of the length $n$ was
calculated from the sum rule Eq.~(\ref{eq:form.5.02}). The probabilistic
treatment of the right hand side of Eq.~(\ref{eq:lvl.03}) is justified for
large enough $n$, when only the zero mode of the classical evolution
operators is important.  This condition can be written as
\begin{equation}
    \tau \gg \tau_\ast\;,
\label{eq:lvl.02}
\end{equation}
where $\tau_\ast = n_\ast / N_H$ is the time of mixing being measured in the 
units of the Heisenberg time. At the moment it is not clear whether we should 
take in to account the correlation between the classical trajectories on the 
Ehrenfest time  scale\cite{Aleiner-Larkin-96}.

The linear in $\tau$ term in the right hand side of Eq.~(\ref{eq:lvl.06})
implies level repulsion and it cannot be correct if $R$ is small and  one
computes the energy levels perturbatively\cite{Berry-apr80}. The perturbation
theory works when $kR<(9\cdot 2^6\pi^2)^{1/8}$, and therefore
Eq.~(\ref{eq:lvl.06}) is valid under the condition $kR\gg 1$. This implies
$\alpha \gg 1$. Since $\alpha\propto \hbar^{-1/3}$ and $\tau_\ast\propto
\hbar$ our result Eq.~(\ref{eq:lvl.06}) demonstrates two regimes
\begin{eqnarray}
    K(\tau) &=& 2\tau -  {N_H\over 2\Omega} \alpha \tau^3
    + O((\alpha\tau)^2)\quad
    \tau_\ast \ll \tau \ll \alpha^{-1}
    \;,
\label{eq:lvl.07}
\\
    K(\tau) &=& 2\tau -  \tau^2 + O(\tau^3)\quad\quad
    \alpha^{-1} \ll \tau \ll 1\;.
\label{eq:lvl.08}
\end{eqnarray}
The last expression matches the universal form-factor $K(\tau) = 2\tau -
\tau^2 + \tau^3/2 + O(\tau^4)$. The universality of the form-factor of 
classically chaotic systems was conjectured\cite{Bohigas-84} on basis of 
numerical data. Our theory supports this conjecture under the condition of
Eq.~(\ref{eq:lvl.08}).

The interference correction to the form-factor, Eq.~(\ref{eq:lvl.07}), can be
obtained on the language of the action correlations\cite{Argaman-dec93}.  One
should pair off the trajectory with two small angle scatterings and the nearby
trajectory without these two scatterings.  Each small angle reflection
changes the phase of the wave by $\pi/3$, then the total phase difference
between two periodic trajectories is $2\pi/3$. This pair contributes $e^{
2\pi i/3}$ to the probability to return.  The time reversal pair contributes
$e^{-2\pi i/3}$ and the sum is just $-1$, that is the right sign of the
interference correction.

The special attention must be paid to the case of the mixed boundary
conditions at the opening of the billiard.\cite{MartinSieber-95} The phase of
the exact scattering matrix is
\begin{eqnarray}
    \phi_l(\kappa) &=& 2\text{arg}\Bigl[
        H_l(kR) - \kappa  H_l'(kR)
    \Bigr] + \pi
\nonumber
\\
    &=&
    \phi_l(0)
    - 2\text{arctan}\Bigl[\kappa\phi_l'(0)/2\Bigr]
\label{eq:mix.2}
\end{eqnarray}
where $H_{l}$ and $H_l'$ are the Hankel function and its derivative, $\kappa$
is the parameter, the degree of the mixing. Since $\phi_l'\approx - {\partial
\phi/\partial l}$ one get
\begin{equation}
   \phi_l(\kappa) \approx \phi_{l+\kappa}(0)\;.
\label{eq:mix.3}
\end{equation}
Therefore the mixed boundary conditions just shift the diffraction edge
leaving the phase at the tangency unchanged. For the large values of $\kappa$
the exact position of the diffraction edge is given by complicated
expression, but the physical mechanism of the interference remains
essentially the same.


\section{ General discussion }

Present work breaks the common believe that the level statistic of the
ergodic systems is determined by symmetry.\cite{Leyvraz-Seligman,AASA-97}
This believe is a result of the analyzes of the effective Lagrangian in
theory of disordered solids\cite{Efetov-jan83}. At the wave number equal to
zero the interaction of the diffusion modes is
universal\cite{ELK-80,Smith-jan98} and implies the universality of the level
statistics.

However, this universality is not occasional, it is the consequence of the
ergodicity and the particle number conservation law. Therefore the
interaction of the Liouvillian modes\cite{AASA-96,ASAA-96} in chaos should be
universal too.  For example one may introduce the $\delta$-correlated
disorder potential as the source of the interference in a chaotic system and
obtain the interaction of the Liouvillian modes independent of the potential
strength.\cite{MK-jul95} The same result may be obtained in the model with
the smooth disorder potential\cite{Aleiner-Larkin-97}.

In this work we consider an example of chaotic system, the Sinai billiard, in
absence of any disorder potential. In such a system, the interference between
classical trajectories takes place because of the hard wall diffraction. In
the semiclassical limit the phase volume of diffraction is relatively small.
The interaction of the Frobenius-Perron modes (we have introduced the area
preserving map instead of the Hamiltonian flow) is proportional to this
volume and it is not universal.  However, the relevant parameter is the time
$\tau_D$, a particle needs to enter the region of diffraction. At low
frequencies $\omega\tau_D\ll1$, or long times $t\gg \tau_D$, the density
evolution operator is not any more Frobenius-Perron, because of diffraction.
According to our results, the interaction between modes of this operator is
universal.

The domain of the universality of the interference effect is, therefore,
$\Delta/\hbar \ll \omega \ll \tau_D^{-1}$ and it depends explicitly on the
phase volume of diffraction. In the case of the Sinai billiard this domain is
large enough and our result supports the universality of the level statistics
observed numerically\cite{Berry-apr80,Bohigas-84}. Generic chaotic system can
have so long $\tau_D$ at given energy, that its level statistics will
never
manifest the universality.

In summary we have found that the hard wall diffraction contributes
the $\tau^2$ and possibly the high order terms into the form-factor of the
energy levels correlation function of the Sinai billiard.

\acknowledgments

Discussions with Prof. U. Smilansky and D. Gutman are gratefully acknowledged.
This work was supported by Israel Science Foundation and the Minerva Center
for Nonlinear Physics of Complex systems.

\end{multicols}


\begin{references}

\bibitem{Smilansky-rev95}
U.~Smilansky,  in {\em Mesoscopic quantum physics}, edited by E.~Akkermans,
  G.~Motambaux, J.-L.~Pichard and J.~Zinn-Justin (Elsevier Science B. V.,
  Amsterdam, 1995), p.\ 373.

\bibitem{Berry-apr80}
M.~V.~Berry, Ann. Phys. {\bf 131},  163  (1981).

\bibitem{Berry-feb85}
M.~V.~Berry,  in {\em Quantum Chaos}, edited by G.~Casati and B.~Chirikov
  (Cambridge university press, Cambridge, 1995), Chap.~Semiclassical theory of
  spectral rigidity, p.\ 319, [Proc. R. Soc. London, {\bf A400}, 229(1985)].

\bibitem{Miller-apr98}
Daniel~L.~Miller, Phys. Rev. E {\bf 57},  4063  (1998), [Preprint
  cond-mat/9705095].

\bibitem{Schuster-book84}
H.~G.~Schuster, {\em Deterministic Chaos} (Physik-Verlag, Weinheim, 1984),
  Chap.~2.

\bibitem{Hannay-OzorioDeAlmeida}
J.~H.~Hannay and A.~M.~Ozorio~De~Almeida, J. Phys. A: Math. Gen. {\bf 17},
  3429  (1984).

\bibitem{Levy-Keller-59}
B.~R.~Levy and J.~B.~Keller, Comm. Pure Appl. Math. {\bf 12},  159  (1959).

\bibitem{Aleiner-Larkin-96}
I.~L.~Aleiner and A.~I.~Larkin, Phys. Rev. B {\bf 54},  14423  (1996).

\bibitem{Hikami-81}
S.~Hikami, Phys. Rev. B {\bf 24},  2671  (1981).

\bibitem{Gorkov-LK-79}
L.~P.~Gor'kov, A.~I.~Larkin and D.~E.~Khmel'nitskii, JETP Lett. {\bf 30},  229
  (1979), [Pis'ma Zh. Eksp. Teor. Fiz. {\bf 30}, 248 (1979)].

\bibitem{Smilansky-rev89}
U.~Smilansky,  in {\em Chaos and quantum physics}, edited by M.-J.~Giannoni,
  A.~Voros, and J.~Zinn-Justin (Elsevier Science B. V., Amsterdam, 1991), p.\
  415.

\bibitem{Berry-87}
M.~V.~Berry, Proc. R. Soc. Lond. A {\bf 413},  183  (1987).

\bibitem{Bohigas-84}
O.~Bohigas, M.~J.~Giannonni, and C.~Schmit, Phys. Rev. Lett. {\bf 52},  1
  (1984).

\bibitem{Argaman-dec93}
N.~Argaman, F.-M.~Dittes, E.~Doron, J.~P.~Keating, A.~Yu.~Kitaev, M.~Sieber,
  and U.~Smilansky, Phys. Rev. Lett. {\bf 71},  4326  (1993).

\bibitem{MartinSieber-95}
M.~Sieber, H.~Primack, U.~Smilansky, I.~Ussishkin and H.~Schanz, J. Phys. A
  {\bf 28},  5041  (1995).

\bibitem{Leyvraz-Seligman}
F.~Leyvraz and T.~H.~Seligman, Phys. Rev. Lett. {\bf 79},  1778  (1997).

\bibitem{AASA-97}
O.~Agam, A.~V.~Andreev, B.~D.~Simons, and B.~L.~Altshuler, Phys. Rev. Lett.
  {\bf 79},  1779  (1997).

\bibitem{Efetov-jan83}
K.~B.~Efetov, Adv. Phys. {\bf 32},  53  (1983).

\bibitem{ELK-80}
K.~B.~Efetov, A.~I.~Larkin, and D.~E.~Khmelnitskii, Sov. Phys. JETP {\bf 52},
  568  (1980), [Zh. Eksp. Teor. Fiz. {\bf 79}, 1120(1980)].

\bibitem{Smith-jan98}
R.~A.~Smith, I.~V.~Lerner, and B.~L.~Altshuler, cond-mat/9801229  .

\bibitem{AASA-96}
A.~V.~Andreev, O.~Agam, B.~D.~Simons, and B.~L.~Altshuler, Phys. Rev. Lett.
  {\bf 76},  3947  (1996).

\bibitem{ASAA-96}
A.~V.~Andreev, B.~D.~Simons, O.~Agam, and B.~L.~Altshuler, Nucl. Phys. B {\bf
  482},  536  (1996), [cond-mat/9605204].

\bibitem{MK-jul95}
B.~A.~Muzykantskii and D.~E.~Khmelnitskii, JETP Lett. {\bf 62},  76  (1995),
  [Pis'ma Zh. Eksp. Teor. Fiz. {\bf 62}, 68(1995)].

\bibitem{Aleiner-Larkin-97}
I.~L.~Aleiner and A.~I.~Larkin, Phys. Rev. E {\bf 55},  1243  (1997).

\end{references}
\end{document}